\documentclass[a4paper,12pt]{article}
\usepackage{amsmath,amssymb,bm}
\usepackage{enumerate}
\usepackage{mathrsfs}
\usepackage{amsthm}

\title{On the Fragility of the Basis on the Hamilton-Jacobi-Bellman Equation in Economic Dynamics}
\author{Yuhki Hosoya\thanks{E-mail: hosoya(at)tamacc.chuo-u.ac.jp}\\Faculty of Economics, Chuo University\thanks{742-1 Higashinakano, Hachioji-shi, Tokyo 192-0393, Japan.}}
\date{\today}
\begin{document}

\maketitle

\begin{abstract}
In this paper, we provide an example of the optimal growth model in which there exist infinitely many solutions to the Hamilton-Jacobi-Bellman equation but the value function does not satisfy this equation. We consider the cause of this phenomenon, and find that the lack of a solution to the original problem is crucial. We show that under several conditions, there exists a solution to the original problem if and only if the value function solves the Hamilton-Jacobi-Bellman equation. Moreover, in this case, the value function is the unique nondecreasing concave solution to the Hamilton-Jacobi-Bellman equation. We also show that without our conditions, this uniqueness result does not hold.

\vspace{12pt}

\vspace{12pt}
\end{abstract}

\section{Introduction}
The Hamilton-Jacobi-Bellman equation (HJB equation) has recently played an important role in the analysis of continuous-time dynamic stochastic general equilibrium (DSGE) models. This equation corresponds to the Bellman equation in discrete-time models, and the value function can be characterized as a solution to this equation. The analysis of this equation allows us to analyze complex problems whose solutions are difficult to evaluate directly. For example, Achdou et al. (2014) is a typical study that used such recent techniques.

However, on closer inspection, we found that there is no literature provides a mathematical foundation for the application of the HJB equation in economic theory. Of course, there are many papers that study the HJB equation itself. However, they are constructed with models used in physics and other fields in mind. Because the structures of models in physics and economics are quite different, it is not possible to apply such studies as they are. The same is true for Pontryagin's maximum principle,\footnote{This technique for variational problems frequently appears in textbooks of macroeconomics. See, for example, Blanchard and Fischer (1989) or Barro and Sala-i-Martin (2003).} but relatively simple methods are available for overcoming this problem in this case. See, for example, Hosoya (2024). However, this method cannot be applied to the HJB equation.

The purpose of this paper is to show the seriousness of this problem. In this paper, we deal not with the DSGE model, but with a much more classical and simple model, called the Ramsey-Cass-Koopmans capital accumulation model.\footnote{This model was originated by Ramsey (1928), and modified by Cass (1965) and Koopmans (1965). Because this model is very common, many textbooks treat it. See, for example, Acemoglu (2009) or Romer (2011).} We then show that, even within the scope of such a simple model, the following examples exist. First, the value function is not a solution to the HJB equation. Generally, the HJB equation is a very difficult class of differential equations, known as degenerate elliptic, and thus it is often the case that there is no ordinary solution (classical solution). However, the value function is not even a `viscosity solution,' which is used in this field as a weak solution.\footnote{This solution concept was introduced by Lions (1982), Crandall and Lions (1983) for the HJB equation.} By contrast, there are infinitely many classical solutions to the HJB equation. In other words, by solving the HJB equation, we obtain myriad candidates for the value function, and in fact, none of them is the true value function (Proposition 1).

Next, we discuss why this problem arose. First, we consider why the usual theory of the HJB equation used in physics cannot be used. There are three possible problems. The first is the difference in time intervals. The second is the difference in the solution concepts. The third is the difference in the objective functions. The first problem means that the models used in economics usually have infinite time intervals, whereas in many applications to physics and other sciences, finite-time problems are considered. The second problem has already been explained. Because there is often no classical solution to the HJB equation, it is usually analyzed using a viscosity solution. However, economists usually need a classical solution. The third problem is that recent studies by mathematicians to overcome the first problem have typically assumed boundedness in the instantaneous utility function. Soner (1986a, b), Baumeister et al. (2007), and Ch.3 of Bardi and Capuzzo-Dolcetta (2008) are typical examples of these studies. However, the most familiar example of the instantaneous utility function for economists is the CRRA function. This class of functions includes the natural logarithmic function, and it is not bounded from above or below. Hence, even the latest mathematical studies on the HJB equation are not directly applicable to economic dynamics.

Actually, Barles (1990) found an example where the HJB equation has a problem when we consider a model with an infinite time interval and unbounded instantaneous utility function. This example suggests that the first and third problems discussed above are essentially important. Despite this, in economics, it is very often discussed as if there were no such examples. One reason, perhaps, is that economists have a `proof' that the value function is a classical solution to the HJB equation. This `proof' is found in classical macroeconomic textbooks. For example, Malliaris and Brock (1982) includes such a `proof.' If this `proof' is correct, then Barles' example is only important for models in fields far from economics and not relevant to economics. However, our Proposition 1 shows that exactly such a problem occurs even in economic models. Hence, this `proof' is actually incorrect.

Then, what problems in the `proof' led to Proposition 1? We examine this and find what appears to be the cause; that is, there is a term that is $o(t)$ if a specific pair of consumption and capital accumulation paths is given, but if we take the supremum over arbitrary consumption and capital accumulation paths, this term may no longer be $o(t)$, and this seems to cause a serious problem.

If a cause is found, it may be possible to remove this cause by making additional assumptions. We expected that evaluating $o(t)$ at the solution without taking a supremum would solve the problem. The result is affirmative. Under weak assumptions, we confirm that the value function satisfies the HJB equation if a solution exists in the original model (Theorem 1).

However, this is far from a solution to the problem we posed in Proposition 1. This is because, in most cases, the existence problem for the solution to continuous-time variational problems cannot be solved by the usual topological methods. See, for example, Ch.9 of Ioffe and Tikhomirov (1979). Thus, there are always difficulties in proving a priori whether a solution exists. We further analyze this and prove the converse of the above result. That is, we prove that under certain conditions, if the value function is a solution to the HJB equation, then we can construct a solution to the original problem using the value function (Theorem 2). Furthermore, in such a case, the value function is the unique nondecreasing and concave solution of the HJB equation (Theorem 3).

These results look good at first glance, but we need to impose some strong conditions on the model. When those conditions are not met, many problems arise. Proposition 1 is such an example. We present another example in which the value function is a solution to the HJB equation, but there are infinitely many other solutions to the HJB equation that are different from the value function (Proposition 2). In this case, the solution can be constructed in the manner of Theorem 2 if the correct value function can be determined, but there is only one value function among myriad solutions to the HJB equation, and it is difficult to determine which solution is the true value function.

We do not view the results of Theorems 2-3 as positive results either. This is for two reasons. First, Theorems 1-2 mean that the difficulty of showing that the value function satisfies the HJB equation is the same as the difficulty of showing directly that a solution to the original problem exists. Because it is difficult to show that a solution exists to the variational problem, it is equally difficult to prove that the value function satisfies the HJB equation. Second, all of our results are presented for deterministic models. As already mentioned, the main application of the HJB equation in economic dynamics is DSGE models, where, because of the stochastic term, the analysis is much more difficult than that for the models we consider. For our problem, it may be possible to show the existence of a solution through the classical analysis of the Euler equations and transversality condition. However, it is not clear whether this can be done with more complex models. In conclusion, we must state that we do not know if the HJB equation can be used with confidence.

The structure of this paper is as follows. In Section 2, we present our most important counterexample: Proposition 1. In Section 3, we examine the logic behind Proposition 1 in more detail. In Section 4, we present some methods to solve the problems that arise there. Because the proofs of theorems are very long, we place them in the appendix. Additionally, we also place the exact definitions of the classical and viscosity solutions to the HJB equation in the appendix. Section 5 is the conclusion.

\section{Example}
Consider the following problem:
\begin{align}
\max~~~~~&~\int_0^{\infty}e^{-\rho t}u(c(t))dt\nonumber \\
\mbox{subject to. }&~k(t)\ge 0,\ c(t)\ge 0,\nonumber \\
&~k(t)\mbox{ is absolutely continuous on any compact set},\nonumber \\
&~c(t)\mbox{ is locally integrable},\label{PROBLEM}\\
&~\int_0^{\infty}e^{-\rho t}u(c(t))dt\mbox{ is defined},\nonumber \\
&~\dot{k}(t)=f(k(t))-c(t),\nonumber \\
&~k(0)=k>0,\nonumber
\end{align}
and corresponding HJB equation:
\begin{equation}\label{HJB}
\sup_{c\ge 0}\{(f(k)-c)V'(k)+u(c)\}=\rho V(k).
\end{equation}
We call a pair of functions $(k(t),c(t))$ from $\mathbb{R}_+$ into $\mathbb{R}_+$ {\bf admissible} if 1) $k(t)$ is absolutely continuous on any compact set, 2) $c(t)$ is locally integrable, 3) $\int_0^{\infty}e^{-\rho t}u(c(t))dt$ is defined, and 4) $\dot{k}(t)=f(k(t))-c(t)$ for almost every $t\in \mathbb{R}_+$.\footnote{Throughout this paper, we use the following notation: $\mathbb{R}_+=\{x\in\mathbb{R}|x\ge 0\}$ and $\mathbb{R}_{++}=\{x\in \mathbb{R}|x>0\}$. Note that the sentence ``$\int_0^{\infty}e^{-\rho t}u(c(t))dt$ is defined'' admits that this value takes $\pm \infty$.} Let $A_k$ be the set of all admissible pairs such that $k(0)=k$. We define the value function of the problem as follows.
\[\bar{V}(k)=\sup\left\{\left.\int_0^{\infty}e^{-\rho t}u(c(t))dt\right|(k(t),c(t))\in A_k\right\}.\]
Now, we show the following result.\footnote{The definition of ``viscosity solution'' is in the appendix.}

\vspace{12pt}
\noindent
{\bf Proposition 1}. Suppose that $\rho>0$, $u(c)=c$, and $f(k)=\sqrt{k}$. Then, the following holds.
\begin{enumerate}[1)]
\item The value function $\bar{V}$ is finite and concave.

\item There are infinitely many classical solutions to the HJB equation.

\item There is no viscosity solution to the HJB equation that is concave.
\end{enumerate}

\vspace{12pt}
\noindent
{\bf Proof}. First, because $u(c)\ge 0$ for every $c\ge 0$, $\int_0^{\infty}e^{-\rho t}u(c(t))dt$ is defined for every nonnegative and measurable function $c(t)$. Therefore, we can omit this requirement.

Define $g(k)=\rho k+\frac{1}{4\rho}$. Because the graph of $g$ is a tangential line to that of $f$ at $k=1/4\rho^2$, $g(k)\ge f(k)$ for every $k\ge 0$. Consider the following modified problem.
\begin{align*}
\max~~~~~&~\int_0^{\infty}e^{-\rho t}u(c(t))dt\\
\mbox{subject to. }&~k(t)\ge 0,\ c(t)\ge 0,\\
&~k(t)\mbox{ is absolutely continuous on any compact set},\\
&~c(t)\mbox{ is locally integrable},\\
&~\dot{k}(t)=g(k(t))-c(t),\\
&~k(0)=k>0.
\end{align*}
Let $B_k$ be the set of all admissible pairs such that $k(0)=k$ in this problem. Define $c^*(t)=\rho k+1/4\rho$ and $k^*(t)=k$. Then, $(k^*(t),c^*(t))\in B_k$, and for every $(k(t),c(t))\in B_k$,
\begin{align*}
\int_0^Te^{-\rho t}(u(c^*(t))-u(c(t)))dt=&~\int_0^Te^{-\rho t}(c^*(t)-c(t))dt\\
=&~\int_0^Te^{-\rho t}[\rho(k^*(t)-k(t))-(\dot{k}^*(t)-\dot{k}(t))]dt\\
=&~\int_0^T\frac{d}{dt}[e^{-\rho t}(k(t)-k^*(t))]dt\\
=&~e^{-\rho T}(k(T)-k^*(T))\\
\ge&~-e^{-\rho T}k^*(T)\to 0\mbox{ as }T\to \infty,
\end{align*}
which implies that $(k^*(t),c^*(t))$ is a solution to the modified problem. On the other hand, suppose that $(k(t),c(t))\in A_k$, and define
\[\tilde{c}(t)=g(k(t))-\dot{k}(t).\]
Because $f(k(t))\le g(k(t))$, we have that $c(t)\le \tilde{c}(t)$ almost everywhere. Clearly, $(k(t),\tilde{c}(t))\in B_k$, and thus $0\le \bar{V}(k)\le k+1/4\rho^2$. Hence, $\bar{V}$ is finite.

Next, let $k_0,k_1>0$, $s\in [0,1]$ and $(k_0(t),c_0(t))\in A_{k_0}, (k_1(t),c_1(t))\in A_{k_1}$, and define
\[k_s=(1-s)k_0+sk_1,\ k_s(t)=(1-s)k_0(t)+sk_1(t),\ c_s(t)=f(k_s(t))-\dot{k}_s(t).\]
By the concavity of $f$,
\[u(c_s(t))=c_s(t)\ge (1-s)c_0(t)+sc_1(t)=(1-s)u(c_0(t))+su(c_1(t)).\]
It is clear that $(k_s(t),c_s(t))\in A_{k_s}$, and thus we have that $\bar{V}$ is concave. Hence, 1) holds.

Next, define
\begin{equation}\label{SOLL}
V(k)=Ae^{2\rho \sqrt{k}-2\rho},
\end{equation}
where $A>0$. Then,
\[V'(k)=\rho V(k)/\sqrt{k}.\]
We can easily verify that
\begin{equation}\label{SOLL2}
\lim_{k\to 0}V'(k)=\lim_{k\to \infty}V'(k)=+\infty,
\end{equation}
which implies that, if $A$ is sufficiently large, then $V'(k)>1$ for every $k>0$. Therefore,
\[\sup_{c\ge 0}\{(f(k)-c)V'(k)+u(c)\}=\sqrt{k}V'(k)=\rho V(k),\]
and thus $V$ is a classical solution to (\ref{HJB}), and 2) holds.

Third, suppose that there is a concave viscosity solution $V$ to (\ref{HJB}). Because every concave function is locally Lipschitz, $V$ is absolutely continuous. This implies that $V$ is differentiable almost everywhere, and if $V$ is differentiable at $k$, then\footnote{Note that, if $V$ is a viscosity solution to (\ref{HJB}) and differentiable at $k$, then $V$ satisfies (\ref{HJB}) at $k$. See subsection A.1 in the appendix.}
\[\sup_{c\ge 0}\{(f(k)-c)V'(k)+u(c)\}=\rho V(k).\]
Because the left-hand side is $+\infty$ if $V'(k)<1$, we have that $V'(k)\ge 1$ for every such $k$. Therefore, we obtain that
\[\sqrt{k}V'(k)=\rho V(k)\]
for every $k$ such that $V$ is differentiable at $k$. By the Caratheodory-Picard-Lindel\"of uniqueness theorem of the solution to ordinary differential equations,\footnote{For the proof of this theorem, see, for example, Ch.1 of Coddington and Levinson (1984) or section 0.4 of Ioffe and Tikhomirov (1979).} we have that $V$ is (\ref{SOLL}), where $A=V(1)$. However, because of (\ref{SOLL2}), $V$ cannot be concave, which is a contradiction. Thus, 3) holds. This completes the proof. $\blacksquare$

\section{Discussion}
In this section, we argue in detail what is occurring in the examples provided in Section 2.

It is easy to create a model such that the value function does not satisfy the HJB equation; that is, because the HJB equation is a differential equation, it is obviously not satisfied for functions that are not real-valued functions. Therefore, if $\bar{V}\equiv +\infty$, then this automatically violates the HJB equation. However, in Proposition 1, the value function of (\ref{PROBLEM}) is finite. This indicates that Proposition 1 presents a more serious example.

By 1) and 3), the value function cannot be a viscosity solution to the HJB equation. Let us consider why this phenomenon has occurred. We convert the explanation of why the value function in the stochastic control problem satisfies the HJB equation in Malliaris and Brock (1982) into the case of this deterministic problem.\footnote{See section 2.10 of Malliaris and Brock (1982). Note that, they considered a problem with a finite time interval, where at the terminal time $N$, the value of $k(N)$ is evaluated by some function $B(N,k(N))$. To apply their argument to our model, we can define $B(N,k)\equiv e^{-\rho N}\bar{V}(k)$. Note also that they assumed that the value function can be affected by the time $t$, and wrote the value function as $J(t,k(t))$. In our model, we can set $J(t,k)=e^{-\rho t}\bar{V}(k)$. These substitutions allow their arguments to be appropriately applied to our problem. In fact, in their section 2.10, the argument when assuming $B(N,k)\equiv 0$ is the most essential (subsection 2.10.1), and the derivation of equation (10.12) in this subsection can be reduced to our argument if we place the assumption that the stochastic term $\sigma$ is $0$.} First, it is easy to show that the following equality holds:
\[\bar{V}(k)=\sup\left\{\left.\int_0^te^{-\rho s}u(c(s))ds+e^{-\rho t}\bar{V}(k(t))\right|(k(s),c(s))\in A_k\right\}.\]
Because $k=k(0)$,
\[\sup\left\{\left.\int_0^t[e^{-\rho s}u(c(s))+\frac{d}{ds}(e^{-\rho s}\bar{V}(k(s)))]ds\right|(k(s),c(s))\in A_k\right\}=0.\]
Using Taylor's theorem, we can modify the above equation into
\[\sup\{[u(c(0))+\bar{V}'(k(0))\dot{k}(0)-\rho \bar{V}(k(0))]t+o(t)|(k(s),c(s))\in A_k\}=0.\]
Because $k(0)=k$ and $\dot{k}(0)=f(k)-c(0)$, for $c=c(0)$,
\begin{equation}\label{EVAL4}
\sup\{[(f(k)-c)\bar{V}'(k)+u(c)-\rho \bar{V}(k)]t+o(t)|(k(s),c(s))\in A_k\}=0.
\end{equation}
Because this function depends only on $c$,
\begin{equation}\label{EVAL5}
\sup_{c\ge 0}\{[(f(k)-c)\bar{V}'(k)+u(c)-\rho \bar{V}(k)]t+o(t)\}=0.
\end{equation}
Dividing by $t$ and letting $t\to 0$, we obtain
\[\sup_{c\ge 0}\{(f(k)-c)\bar{V}'(k)+u(c)-\rho \bar{V}(k)\}=0,\]
which is equivalent to (\ref{HJB}).

If the above `proof' is correct, then the value function must be a classical solution to the HJB equation. However, in practice, as we observed in Proposition 1, the value function may not even be a viscosity solution to the HJB equation. Therefore, the above `proof' is wrong. Although the above `proof' includes many gaps that can cause the error,\footnote{For example, in the above `proof', we assume that $\bar{V}(k)$ is differentiable, $k(t)$ is differentiable, and $\int_0^te^{-\rho s}u(c(s))ds$ is differentiable. These may not hold when we do not assume anything on $(k(t),c(t))$. However, we think that these problems are not significant. Note also that, we can easily confirm that these assumptions automatically hold in the model (\ref{PROBLEM2}).} we think that a problem concerning the evaluation of $o(t)$ is the most serious. As seen above, up to (\ref{EVAL4}), the supremum over $(k(s),c(s))\in A_k$ is calculated, but in (\ref{EVAL5}), the supremum over $c\ge 0$ is used. We stated that the reason for this is that the function in (\ref{EVAL4}) only depends on $c$. However, in fact, $o(t)$ depends not only on $c$ but also on the entire trajectory of $(k(s),c(s))$. Moreover, it is not certain whether this part, which could have been $o(t)$ if $(k(s),c(s))$ were fixed, is also $o(t)$ when the supremum is taken over $(k(t),c(t))$.

In (\ref{PROBLEM}), we admit the discontinuity of $c(t)$. However, Proposition 1 still holds even if we consider the following problem:
\begin{align}
\max~~~~~&~\int_0^{\infty}e^{-\rho t}u(c(t))dt\nonumber \\
\mbox{subject to. }&~k(t)\ge 0,\ c(t)\ge 0,\nonumber \\
&~k(t)\mbox{ is continuously differentiable},\nonumber \\
&~c(t)\mbox{ is continuous},\label{PROBLEM2}\\
&~\int_0^{\infty}e^{-\rho t}u(c(t))dt\mbox{ is defined},\nonumber \\
&~\dot{k}(t)=f(k(t))-c(t),\nonumber \\
&~k(0)=k>0,\nonumber
\end{align}
because we do not use the discontinuity of $c(t)$ in the proof of Proposition 1. Many gaps in the above `proof' vanish in (\ref{PROBLEM2}), but the gap that concerns `$o(t)$' does not vanish. Hence, we believe that this problem is truly the problem in the above `proof.' In other words, the above `proof' collapsed because of an overly cursory evaluation of $o(t)$.

We consider this point in more detail in the next section.

\section{Equivalence Result for the Existence of Solutions}
In this section, we still consider the problem (\ref{PROBLEM}). However, the discontinuity in $c(t)$ does not become important, and thus if the reader considers (\ref{PROBLEM2}) to be more favorable, he/she may continue reading with that in mind. Some changes to the proof are required, but they do not pose a major problem.\footnote{See subsection B.1.}

Because we want to build a general theory, we introduce some assumptions on $\rho$, $u$, and $f$.

\vspace{12pt}
\noindent
{\bf Assumption R}. $\rho>0$.

\vspace{12pt}
\noindent
{\bf Assumption U}. $u:\mathbb{R}_+\to \mathbb{R}\cup \{-\infty\}$ is continuous, concave, and increasing on $\mathbb{R}_+$, and continuously differentiable on $\mathbb{R}_{++}$.\footnote{Note that, because $u$ is increasing, $u(c)\in \mathbb{R}$ if $c>0$.}

\vspace{12pt}
\noindent
{\bf Assumption F}. $f:\mathbb{R}_+\to \mathbb{R}$ is continuous and concave. Moreover, $f(0)=0$.

\vspace{12pt}
We call an admissible pair $(k^*(t),c^*(t))\in A_k$ a {\bf solution} to (\ref{PROBLEM}) if the following two requirements hold. First,
\[\int_0^{\infty}e^{-\rho t}u(c^*(t))dt\in \mathbb{R}.\]
Second, if $(k(t),c(t))\in A_k$, then
\[\int_0^{\infty}e^{-\rho t}u(c^*(t))dt\ge \int_0^{\infty}e^{-\rho t}u(c(t))dt.\]
Note that, by the first requirement, we do not say that $(k^*(t),c^*(t))$ is a solution if
\[\int_0^{\infty}e^{-\rho t}u(c^*(t))dt=+\infty.\]
The above two requirements can be summarized as
\[\int_0^{\infty}e^{-\rho t}u(c^*(t))dt=\bar{V}(k)\in \mathbb{R}.\]
Our first theorem is as follows.

\vspace{12pt}
\noindent
{\bf Theorem 1}. Suppose that Assumptions R, U, and F hold. Moreover, suppose that for every $k>0$, there exists a solution $(k^*(t),c^*(t))$ to (\ref{PROBLEM}) such that $c^*(t)$ is continuous. Then, the value function $\bar{V}$ is a classical solution to (\ref{HJB}).

\vspace{12pt}
Therefore, our problem vanishes if the model (\ref{PROBLEM}) has a solution.

However, we do not consider that the problem indicated by Proposition 1 is solved by Theorem 1. First, problems such as (\ref{PROBLEM}) are difficult to prove directly that a solution exists. Unlike discrete-time problems, the set of admissible functions in continuous-time models is not compact in the usual topology, and conversely, the objective function is not continuous in the topology where this set becomes compact. Thus, it is necessary to show the existence and characterization of the solution using other methods. (\ref{HJB}) is seen as a powerful tool for this. However, Theorem 1 states that the value function satisfies (\ref{HJB}) `if a solution exists', which does not solve the problem.

It is interesting that the converse of Theorem 1 can be proved by slightly strengthening the assumptions.

\vspace{12pt}
\noindent
{\bf Theorem 2}. Suppose that Assumptions R, U, and F hold, and the following requirements hold.
\begin{enumerate}[(i)]
\item The function $u'$ is decreasing and $u'(\mathbb{R}_{++})=\mathbb{R}_{++}$.

\item There exist $k_1,k_2>0$ and $p_2\in \partial f(k_2)$ such that $D_+f(k_1)>\rho>p_2>0$, where $D_+f$ denotes the right-side derivative of $f$ and $\partial f$ denotes the subdifferential of $f$.\footnote{See subsection A.2 in the appendix.}
\end{enumerate}
Moreover, suppose that the value function of (\ref{PROBLEM}) is a classical solution to (\ref{HJB}). Consider the following ordinary differential equation:
\begin{equation}\label{SOL}
\dot{k}(t)=f(k(t))-(u')^{-1}(\bar{V}'(k(t))),\ k(0)=k.
\end{equation}
Then, there exists a solution $k^*(t)$ to (\ref{SOL}) defined on $\mathbb{R}_+$. Furthermore, if we define $c^*(t)=(u')^{-1}(\bar{V}'(k^*(t)))$, then $(k^*(t),c^*(t))$ is a solution to (\ref{PROBLEM}).

\vspace{12pt}
As a condition for computing the value function by solving (\ref{HJB}), the following theorem provides an interesting suggestion.

\vspace{12pt}
\noindent
{\bf Theorem 3}. Suppose that Assumptions R, U, and F hold, and (i) and (ii) of Theorem 2 also hold. Moreover, suppose that for every $k>0$, there exists a solution $(k^*(t),c^*(t))$ to (\ref{PROBLEM}) such that $c^*(t)$ is continuous. Then, $\bar{V}$ is the unique concave increasing classical solution to (\ref{HJB}).

\vspace{12pt}
Therefore, if assumptions (i) and (ii) are met, and there exists a solution to (\ref{PROBLEM}), then we can obtain the value function by solving (\ref{HJB}), and then a solution to (\ref{PROBLEM}) by solving (\ref{SOL}).

However, we consider that this result is not positive. According to Theorems 1-2, under slightly stronger assumptions, the value function satisfies (\ref{HJB}) if and only if there exists a solution to (\ref{PROBLEM}). In other words, the difficulty of showing that the value function satisfies (\ref{HJB}) is equivalent to that of showing the existence of a solution. This fundamentally indicates the difficulty that we have in showing the existence of a solution using the value function.

The additional assumptions (i) and (ii) for $u$ and $f$ given in Theorem 2 are important. In fact, if we solve (\ref{HJB}) in a model where these requirements do not hold and derive a candidate of the value function $V(k)$, and then solve (\ref{SOL}) using this instead of $\bar{V}$, it may not be a solution to (\ref{PROBLEM}). In the case of Proposition 1, this is obvious because the value function is not a solution to (\ref{HJB}). However, even if the value function solves (\ref{HJB}) and there always exists a solution to the original problem, there are still cases in which this cannot be used. A specific example is provided by the following proposition.

\vspace{12pt}
\noindent
{\bf Proposition 2}. Consider (\ref{PROBLEM}), where $\rho=1, u(c)=c+\sqrt{c}$, and $f(k)=k$. Then, the value function $\bar{V}(k)=k+\sqrt{k}$ satisfies (\ref{HJB}), and for each $k>0$, there is a solution $(k^*(t),c^*(t))$ to (\ref{PROBLEM}). However, there exist infinitely many solutions to (\ref{HJB}) other than $\bar{V}(k)$.

\vspace{12pt}
\noindent
{\bf Proof}. First, let $k^*(t)=k$ and $c^*(t)=k$. Then, this pair $(k^*(t),c^*(t))$ satisfies the Euler equation and transversality condition, and thus this is a solution to (\ref{PROBLEM}).\footnote{For a proof, see Theorem 2 of Hosoya (2024).} This means that $\bar{V}(k)=k+\sqrt{k}$. Next, (\ref{HJB}) can be transformed as follows:
\[kV'(k)+\frac{1}{4(V'(k)-1)}=V(k).\]
This is a sort of Clairaut's equation, where the general solution is
\[Ak+\frac{1}{4(A-1)},\]
where $A>1$. On the other hand, the singular solution to this equation is $k+\sqrt{k}$, which coincides with the value function $\bar{V}$. Therefore, the value function satisfies (\ref{HJB}), but there are many other solutions. This completes the proof. $\blacksquare$

\vspace{12pt}
Suppose that one solves (\ref{HJB}) for this problem, derives the general solution $Ak+\frac{1}{4(A-1)}$, and then solves (\ref{SOL}), assuming that this is the value function. Then (\ref{SOL}) becomes the following equation.
\[\dot{k}(t)=k(t)-\frac{1}{4(A-1)^2}.\]
Of course, solving this does not provide a solution to the problem, because either $k^*(t)$ cannot be positive on $\mathbb{R}_+$ or the transversality condition is not satisfied whenever $k\neq \frac{1}{4(A-1)^2}$. In other words, the technique of solving (\ref{HJB}) and using (\ref{SOL}) to derive the solution cannot be used in this case.

It should be noted that, in the proof of Proposition 2, we first obtain a solution and use it to calculate the value function. In fact, to determine the value function directly, we must find a solution. Therefore, when there are many solutions to (\ref{HJB}), we cannot determine which of them is the real value function without solving the original problem. Of course, there are cases like Proposition 1, where there are many solutions, but none of them are value functions. Therefore, this method breaks down.

The Euler equation and transversality condition, known as a sufficient condition for a solution, have a certain role in solving this problem. That is, if we can say that there exists a pair that satisfies the Euler equation and transversality condition, then it is a solution, and therefore a solution exists. By Theorem 1, at least the value function satisfies (\ref{HJB}). But what role can (\ref{HJB}) have in that case? There is no need to ensure the existence of a solution, since we have already said that the solution exists. If we want to use it to compute the solution, then we encounter the problem posed in Proposition 2. That is, we do not know which solution to (\ref{HJB}) is the real value function.

Theorem 3 indicates that under conditions (i) and (ii) in Theorem 2, nothing like Proposition 2 can occur. At first sight, this appears to solve the problem. However, this occurred because (\ref{PROBLEM}) is very simple. It is not at all clear under what conditions a result like Theorem 3 would hold in a more complex model in which, for example, stochastic shocks are introduced. And (\ref{HJB}) is most actively used in such models. This is, in our view, a serious problem.

Finally, it is worth mentioning Theorem 2 of Hosoya (2023). Using this result, we can easily confirm that the value function is the unique classical solution to (\ref{HJB}) on the space of all concave, increasing functions $V$ that satisfies a growth condition (\ref{GC2}) (see the appendix) if 1) there exists $a>0,b\in\mathbb{R},\theta>0$ such that $u(c)\le au_{\theta}(c)+b$ for all $c\ge 0$, where $u_{\theta}$ denotes a CRRA function with parameter $\theta$,\footnote{That is,
\[u_{\theta}(c)=\begin{cases}
\frac{c^{1-\theta}-1}{1-\theta} & \mbox{if }\theta\neq 1,\\
\log c & \mbox{if }\theta=1.
\end{cases}\]
This function is the unique solution to the following differential equation:
\[-\frac{cu''(c)}{u'(c)}=\theta,\ u(1)=0,\ u'(1)=1,\]
where the left-hand side is sometimes called the {\bf relative risk aversion} of the function $u$. Therefore, $u_{\theta}$ is called the constant relative risk aversion function, and abbreviatedly, the CRRA function.} and 2) there exist $k_1, k_2>0$ and $p_2\in \partial f(k_2)$ such that $D_+f(k_1)>\rho$, $p_2>0$, and $\rho-(1-\theta)p_2>0$. However, the proof of this fact is very difficult, and it is not known whether this proof can extend to the case of the DSGE model.

\section{Conclusion}
We argued on the HJB equation in economic dynamics, and found an example such that the value function is not a viscosity solution to the HJB equation, although there are infinitely many classical solutions to the HJB equation. We analyzed why this example arose, and found that the lack of a solution is crucial. We showed that if the solution to the original problem exists, then the value function is always a solution to the HJB equation. Moreover, if additional assumptions hold, then the value function solves the HJB equation if and only if there always exists a solution to the original variational problem, and in this case, the value function is actually the unique nondecreasing concave solution to the HJB equation. Finally, we provided another example in which the additional assumption is violated, and there are infinitely many nondecreasing concave solutions to the HJB equation, although the value function itself is included in.

We believe that, in this paper, we have adequately demonstrated the difficulty of dealing with the HJB equation in economic dynamics. Of course, we do not intend to argue that the HJB equation should not be used in economic dynamics. In fact, we can compute the solution using this equation and (\ref{SOL}) when some additional conditions are satisfied. It is not impossible to make the same argument for more complex models. Theorem 2 of Hosoya (2023a) is an example of such results. We would like to emphasize that in order to use the HJB equation, one must have the proper ancillary conditions, and when these are not satisfied, many strange counterexamples will appear. Therefore, we must develop the mathematical basis to safely handle this equation in more complex models, such as the DSGE model. We believe that this is a task that must be done as soon as possible.

\appendix
\section{Preliminary}
\subsection{Knowledge on the HJB equation}
Recall the HJB equation (\ref{HJB}):\footnote{Note that, our HJB equation does not include any boundary condition. In this context, there are two types of HJB equations: with and without boundary conditions. Moreover, boundary conditions include several types. For example, Crandall and Lions (1983) and Ishii (1989) argued on an HJB equation in which the boundary values were specified. Soner (1986a,b) did not specify the boundary values of the solution a priori; instead, he treated an endogenous boundary condition. Baumeister et al. (2007), Hermosilla and Zidani (2015), and Hermosilla et al. (2017) considered HJB equations without boundary conditions. Ch.11 of \"Oksendal (2010) treated an HJB equation with stochastic shocks, where the exit time $T$ may be finite. Because the exit time may become finite in this model, there must be a sort of boundary condition. Ch.3 of Bardi and Capuzzo-Dolcetta (2008) did not treat any boundary condition, whereas Ch.4 treated both the Soner--type and the \"Oksendal--type boundary conditions. Thus, some books and papers introduce boundary conditions when dealing with the HJB equation, while others do not. In the context of macroeconomic dynamics, a boundary condition is usually not treated, because the natural restriction to the number $\bar{V}(0)$ is absent in the model.}
\[\sup_{c\ge 0}\{(f(k)-c)V'(k)+u(c)\}=\rho V(k).\]
A continuously differentiable function $V:\mathbb{R}_{++}\to \mathbb{R}$ is called a \textbf{classical solution} to the HJB equation if and only if equation (\ref{HJB}) holds for every $k>0$.

It is known that, in many models of the optimal control problem, there exists no classical solution to the HJB equation. Hence, we should extend the notion of the solution. First, an upper semi-continuous function $V:\mathbb{R}_{++}\to \mathbb{R}$ is called a \textbf{viscosity subsolution} to (\ref{HJB}) if and only if for every $k>0$ and every continuously differentiable function $\varphi$ defined on a neighborhood of $k$ such that $\varphi(k)=V(k)$ and $\varphi(k')\le V(k')$ whenever both sides are defined,
\[\sup_{c\ge 0}\{(f(k)-c)\varphi'(k)+u(c)\}\le \rho V(k).\]
Second, a lower semi-continuous function $V:\mathbb{R}_{++}\to \mathbb{R}$ is called a \textbf{viscosity supersolution} to (\ref{HJB}) if and only if for every $k>0$ and every continuously differentiable function $\varphi$ defined on a neighborhood of $k$ such that $\varphi(k)=V(k)$ and $\varphi(k')\ge V(k')$ whenever both sides are defined,
\[\sup_{c\ge 0}\{(f(k)-c)\varphi'(k)+u(c)\}\ge \rho V(k).\]
If a continuous function $V:\mathbb{R}_{++}\to \mathbb{R}$ is both viscosity sub- and supersolution to (\ref{HJB}), then $V$ is called a \textbf{viscosity solution} to (\ref{HJB}).\footnote{Usually, the HJB equation is discussed in the context of minimization problems, although the problems addressed in this paper are maximization problems. As a result, the inequalities on $\varphi$ appearing in the above definitions are given in the opposite direction from the usual cases.}

Suppose that $V$ is a viscosity solution to the HJB equation and is differentiable at $k>0$. Then, it is known that
\[\sup_{c\ge 0}\{(f(k)-c)V'(k)+u(c)\}=\rho V(k).\]
For a proof, see Proposition 1.9 of Ch.2 of Bardi and Capuzzo-Dolcetta (2008).

\subsection{Subdifferentials and Left- and Right-Derivatives}
In the proof of Theorem 2, we need several results for subdifferentials of the concave function. Hence, in this subsection, we introduce the notion of the subdifferential and needed results. For the proofs of these results, see textbooks on convex analysis, such as Rockafeller (1996).

Suppose that a function $G:U\to \mathbb{R}$ is concave, $U\subset \mathbb{R}^n$ is convex, and the interior $V$ of $U$ is nonempty. Choose any $x\in V$. Define
\[\partial G(x)=\{p\in \mathbb{R}^n|G(y)-G(x)\le p\cdot (y-x)\mbox{ for all }y\in U\}.\]
Then, we can show that $\partial G(x)$ is nonempty. The set-valued mapping $\partial G$ is called the \textbf{subdifferential} of $G$.\footnote{Formally, the subdifferential is defined for not concave but \textbf{convex} functions, and thus the inequality in the definition is reversed. In this view, the name `subdifferential' may not be appropriate, and `superdifferential' may be more suitable. However, in the literature of economics, these two notions are not distinguished, and thus our $\partial G$ is traditionally called the `subdifferential'.}

If $n=1$, then define the left- and right-derivatives $D_-G(x),\ D_+G(x)$ such as
\[D_-G(x)=\lim_{y\uparrow x}\frac{G(y)-G(x)}{y-x},\ D_+G(x)=\lim_{y\downarrow x}\frac{G(y)-G(x)}{y-x}.\]
Note that, if $G$ is concave, then $\frac{G(y)-G(x)}{y-x}$ is nonincreasing in $y$, and thus
\[D_-G(x)=\inf_{t>0}\frac{G(x-t)-G(x)}{-t},\ D_+G(x)=\sup_{t>0}\frac{G(x+t)-G(x)}{t},\]
which implies that both $D_-G(x),D_+G(x)$ are defined and real numbers on $V$. It is known that $\partial G(x)=[D_+G(x),D_-G(x)]$.

Suppose that $x_1<x_2$, $G$ is a concave function defined on an open interval including $[x_1,x_2]$, $p\ge r\ge q, p\in \partial G(x_1)$, and  $q\in \partial G(x_2)$. We show that there exists $x\in [x_1,x_2]$ such that $r\in \partial G(x)$. If $r\ge D_+G(x_1)$, then $r\in \partial G(x_1)$. If $r\le D_-G(x_2)$, then $r\in \partial G(x_2)$. Therefore, we assume that $D_-G(x_2)<r<D_+G(x_1)$. Define $g(k)=G(x)-rx$. Then, $D_+g(x_1)>0$ and $D_-g(x_2)<0$, and thus, there exists $x\in ]x_1,x_2[$ such that $g(x)=\max_{y\in [x_1,x_2]}g(y)$.\footnote{Note that, any concave function on an open interval is continuous, and thus $G$ is continuous on $[x_1,x_2]$.} By the definition of the subdifferential, $0\in \partial g(x)$, and thus $r\in \partial G(x)$, as desired. In particular, if we set $p=D_+G(x_1)$, $q=D_-G(x_2)$, and $r=\frac{G(x_2)-G(x_1)}{x_2-x_1}$, we obtain the {\bf mean value theorem} for subdifferentials: that is, if $G$ is a concave function defined on an open set including $I=[x_1,x_2]$, then there exists $x\in I$ such that $\frac{G(x_2)-G(x_1)}{x_2-x_1}\in \partial G(x)$. We use this result in the proof of Theorem 2.

\section{Proof of Theorem 1}
\subsection{Preparation}
Suppose that the value function $\bar{V}(k)$ is finite for all $k>0$, and fix $k>0$. In the proof of Theorem 1, we frequently use the following evaluations. First, choose any $(k(s),c(s))\in A_k$. For $t>0$, if $k(t)>0$, then
\begin{equation}\label{EVAL6}
\bar{V}(k)\ge \int_0^te^{-\rho s}u(c(s))ds+e^{-\rho t}\bar{V}(k(t)).
\end{equation}
Second, if $(k^*(s),c^*(s))$ is a solution to (\ref{PROBLEM}) and $k^*(t)>0$ for some $t>0$, then
\begin{equation}\label{EVAL7}
\bar{V}(k)=\int_0^te^{-\rho s}u(c^*(s))ds+e^{-\rho t}\bar{V}(k^*(t)).
\end{equation}
In this subsection, we derive these evaluations.

First, let $(k(s),c(s))\in A_k$, $t>0$, and $k(t)>0$. Choose any $\varepsilon>0$. Then, there exists $(\bar{k}(s),\bar{c}(s))\in A_{k(t)}$ such that
\[\int_0^{\infty}e^{-\rho s}u(\bar{c}(s))ds>\bar{V}(k(t))-\varepsilon.\]
Define
\[(\hat{k}(s),\hat{c}(s))=\begin{cases}
(k(s),c(s)) & \mbox{if }0\le s\le t,\\
(\bar{k}(s-t),\bar{c}(s-t)) & \mbox{if }s>t.
\end{cases}\]
Then, $(\hat{k}(s),\hat{c}(s))\in A_k$, and thus
\begin{align*}
&~\int_0^te^{-\rho s}u(c(s))ds+e^{-\rho t}\bar{V}(k(t))\\
<&~\int_0^{\infty}e^{-\rho s}u(\hat{c}(s))ds+e^{-\rho t}\varepsilon\\
\le&~\bar{V}(k)+\varepsilon,
\end{align*}
which implies that (\ref{EVAL6}) holds.

Second, let $(k^*(s),c^*(s))\in A_k$ be a solution to (\ref{PROBLEM}) and $k^*(t)>0$ for some $t>0$. Define $(k^+(s),c^+(s))=(k^*(s+t),c^*(s+t))$. Then, $(k^+(s),c^+(s))\in A_{k^*(t)}$. If $(k^+(s),c^+(s))$ is not a solution to (1) when $k$ is replaced with $k^*(t)$, then there exists $(\bar{k}(s),\bar{c}(s))\in A_{k^*(t)}$ such that
\[\int_0^{\infty}e^{-\rho s}u(c^+(s))ds<\int_0^{\infty}e^{-\rho s}u(\bar{c}(s))ds.\]
Define
\[(\hat{k}(s),\hat{c}(s))=\begin{cases}
(k^*(s),c^*(s)) & \mbox{if }0\le s\le t,\\
(\bar{k}(s-t),\bar{c}(s-t)) & \mbox{if }s>t.
\end{cases}\]
Then, $(\hat{k}(s),\hat{c}(s))\in A_k$ and
\[\int_0^{\infty}e^{-\rho s}u(\hat{c}(s))ds>\int_0^{\infty}e^{-\rho s}u(c^*(s))ds,\]
which contradicts that $(k^*(t),c^*(t))$ is a solution. Therefore,
\begin{align*}
e^{-\rho t}\bar{V}(k^*(t))=&~\int_0^{\infty}e^{-\rho (s+t)}u(c^+(s))ds\\
=&~\int_t^{\infty}e^{-\rho s}u(c^*(s))ds,
\end{align*}
which implies that (\ref{EVAL7}) holds.

Suppose that we consider the problem (\ref{PROBLEM2}) instead of (\ref{PROBLEM}). Then, in both derivations, $\hat{c}(s)$ may be discontinuous, and thus $(\hat{k}(s),\hat{c}(s))$ may be not admissible. However, $\hat{c}(s)$ is still piecewise continuous, and any piecewise continuous function can be approximated by a continuous function from below. Therefore, we can derive the same evaluations in the same manner. This is the only change in the proof when we consider (\ref{PROBLEM2}) instead of (\ref{PROBLEM}).

\subsection{Proof of Theorem 1}
We can easily check that this problem (\ref{PROBLEM}) satisfies all the requirements of Theorem 2 in Benveniste and Scheinkman (1979), and thus the value function $\bar{V}$ is continuously differentiable. Choose any $k_0,k_1>0$. Suppose that $(k_0(t),c_0(t))\in A_{k_0}$ and $(k_1(t),c_1(t))\in A_{k_1}$. For $s\in [0,1]$, define
\[k_s=(1-s)k_0+sk_1,\ k_s(t)=(1-s)k_0(t)+k_1(t),\ c_s(t)=f(k_s(t))-\dot{k}_s(t).\]
Because $f$ is concave,
\[c_s(t)\ge (1-s)c_0(t)+sc_1(t),\]
and $(k_s(t),c_s(t))\in A_{k_s}$. This implies that $\bar{V}$ is concave.

Choose any $c>0$, and consider the following differential equation:
\[\dot{k}(t)=f(k(t))-c,\ k(0)=k.\]
By the Picard-Lindel\"of existence theorem, there exists $t^*>0$ and a positive continuously differentiable solution $k(t)$ to the above equation defined on $[0,t^*]$. By (\ref{EVAL6}), if $0<t\le t^*$,
\[\bar{V}(k)\ge \int_0^te^{-\rho s}u(c)ds+e^{-\rho t}\bar{V}(k(t)).\]
This implies that
\[\frac{e^{-\rho t}\bar{V}(k(t))-e^{-\rho 0}\bar{V}(k(0))}{t}\le -\frac{1}{t}\int_0^te^{-\rho s}u(c)ds.\]
Letting $t\to 0$, we have that
\[-\rho \bar{V}(k)+\bar{V}'(k)(f(k)-c)\le -u(c),\]
and thus,
\[(f(k)-c)\bar{V}'(k)+u(c)\le \rho \bar{V}(k).\]
Because the left-hand side is continuous in $c$,
\[\sup_{c\ge 0}\{(f(k)-c)\bar{V}'(k)+u(c)\}\le \rho \bar{V}(k).\]
To prove the converse inequality, choose a solution $(k^*(t),c^*(t))\in A_k$ such that $c^*(t)$ is continuous. By (\ref{EVAL7}), for any small $t>0$ such that $k^*(t)>0$,
\[\bar{V}(k)=\int_0^te^{-\rho s}u(c^*(s))ds+e^{-\rho t}\bar{V}(k^*(t)).\]
Because $\bar{V}(k)$ is concave,
\begin{align*}
\int_0^t(c^*(s)-f(k^*(s)))\bar{V}'(k)ds=&~\bar{V}'(k)(k-k^*(t))\\
\le&~\bar{V}(k)-\bar{V}(k^*(t))\\
=&~\int_0^te^{-\rho s}u(c^*(s))ds+(e^{-\rho t}-1)\bar{V}(k^*(t)).
\end{align*}
Dividing both sides by $t$ and letting $t\to 0$, we obtain
\[-(f(k)-c^*(0))\bar{V}'(k)\le u(c^*(0))-\rho \bar{V}(k),\]
and thus,
\begin{align*}
\rho \bar{V}(k)\le&~(f(k)-c^*(0))\bar{V}'(k)+u(c^*(0))\\
\le&~\sup_{c\ge 0}\{(f(k)-c)\bar{V}'(k)+u(c)\},
\end{align*}
which implies that
\[\sup_{c\ge 0}\{(f(k)-c)\bar{V}'(k)+u(c)\}=\rho \bar{V}(k),\]
as desired. This completes the proof. $\blacksquare$

\section{Proofs of Theorems 2 and 3}
\subsection{Global Existence of the Pure Accumulation Path}
First, we consider the following differential equation:
\begin{equation}
\dot{k}(t)=f(k(t)),\ k(0)=k,\label{NC}
\end{equation}
and let $\hat{k}(t,k)$ be the function such that for fixed $k>0$, $t\mapsto \hat{k}(t,k)$ denotes the nonextendable solution to the above equation. This function $\hat{k}(t,k)$ is called a {\bf pure accumulation path}. In this subsection, we show that $\hat{k}(t,k)$ is defined on $\mathbb{R}_+\times \mathbb{R}_{++}$, and $\inf_{t\ge 0}\hat{k}(t,k)>0$ under Assumption F and (ii) of Theorem 2.

We first introduce a lemma.\footnote{This lemma is a simplified result of Theorem 1 of Hosoya (2019).}

\vspace{12pt}
\noindent
{\bf Lemma 1}. Suppose that the real-valued functions $h_1(t,k),h_2(t,k)$ defined on $\mathbb{R}_+\times \mathbb{R}_{++}$ are continuous in $k$, measurable in $t$, and satisfy $h_1(t,k)\le h_2(t,k)$ for all $(t,k)\in \mathbb{R}_+\times \mathbb{R}_{++}$. Suppose also that there exists $i^*\in \{1,2\}$ such that $h_{i^*}$ is continuous in $(t,k)$ and locally Lipschitz in $k$.\footnote{That is, for every $(t,k)\in \mathbb{R}_+\times \mathbb{R}_{++}$, there exists a neighborhood $U\subset \mathbb{R}_+\times \mathbb{R}_{++}$ of $(t,k)$ and $L>0$ such that if $(t,k_1), (t,k_2)\in U$, then
\[|h_{i^*}(t,k_1)-h_{i^*}(t,k_2)|\le L|k_1-k_2|.\]} Consider the following differential equations:
\[\dot{k}_i(t)=h_i(t,k_i(t)),\ k_i(0)=k_i^*,\]
where $k_2^*\ge k_1^*>0$, and suppose that there are solutions $k_i:[0,T]\to \mathbb{R}_{++}$ to the above equation for $i\in \{1,2\}$. Then, $k_1(t)\le k_2(t)$ for all $t\in [0,T]$.\footnote{If $i^*$ is absent, then we can find a counterexample. For example, choose $h_1(t,k)=\sqrt{|k-1|}-\frac{t}{8}$, $h_2(t,k)=\sqrt{|k-1|}$, $k_1^*=k_2^*=1$, $k_1(t)=1+\frac{t^2}{16}$, and $k_2(t)=1$.}

\vspace{12pt}
\noindent
{\bf Proof of Lemma 1}. We treat only the case $i^*=2$, since the remaining case can be symmetrically treated. 

Suppose not. Then, there exists $t^*\in [0,T]$ such that $k_1(t^*)>k_2(t^*)$. Because $k_1(0)=k_1^*\le k_2^*=k_2(0)$, we have that $t^*>0$. Define $t^+=\inf\{t\in [0,t^*]|k_1(s)>k_2(s)\mbox{ for all }s\in [t,t^*]\}$. Because $k_1(0)=k_1^*\le k_2^*=k_2(0)$, we have that $k_1(t^+)=k_2(t^+)$. Define
\[k_3(t)=k_1(t^+)+\int_{t^+}^th_2(s,k_1(s))ds.\]
Then, $k_3(t)$ is defined on $[t^+,t^*]$. Moreover,
\[k_3(t)\ge k_1(t^+)+\int_{t^+}^th_1(s,k_1(s))ds=k_1(t)\ge k_2(t)\]
for every $t\in [t^+,t^*]$. Choose a neighborhood $U$ of $(t^+,k_1(t^+))$ and $L>0$ such that if $(t,k_1),(t,k_2)\in U$, then
\[|h_2(t,k_1)-h_2(t,k_2)|\le L|k_1-k_2|.\]
We can assume without loss of generality that $(t,k_1(t)),(t,k_2(t))\in U$ for all $t\in [t^+,t^*]$. Thus,
\begin{align*}
k_3(t)-k_2(t)=&~\int_{t^+}^t[h_2(s,k_1(s))-h_2(s,k_2(s))]ds\\
\le&~\int_{t^+}^tL[k_1(s)-k_2(s)]ds\\
\le&~L(t-t^+)\max_{s\in [t^+,t]}(k_1(s)-k_2(s)).
\end{align*}
Fix some $t\in [t^+,t^*]$ with $0<t-t^+<L^{-1}$, and let $s^*\in \arg\max\{k_1(s)-k_2(s)|s\in [t^+,t]\}$. Because $k_1(t^+)=k_2(t^+)$ and $k_1(s)>k_2(s)$ for $s\in ]t^+,t^*]$, $s^*>t^+$. Thus,
\[k_3(s^*)-k_2(s^*)\le L(s^*-t^+)(k_1(s^*)-k_2(s^*))<k_1(s^*)-k_2(s^*),\]
and hence $k_3(s^*)<k_1(s^*)$, which is a contradiction. This completes the proof of Lemma 1. $\blacksquare$

\vspace{12pt}
Fix $k>0$. Because $f$ is concave, it is locally Lipschitz on $\mathbb{R}_{++}$. By the Picard-Lindel\"of existence theorem, there exists a solution to (\ref{NC}) defined on $[0,T]$. Suppose that the positive nonextendable solution $k^*(t)$ is defined only on $[0,t^*[$, where $t^*<\infty$.\footnote{We call a solution $k^*(t)$ to (\ref{NC}) {\bf positive nonextendable} if $k^*(t)>0$ for all $t$ and there is no positive solution $k(t)$ to (\ref{NC}) such that $k(t)=k^*(t)$ if both sides are defined and the domain of $k(t)$ is wider than that of $k^*(t)$.} By Theorem 3.1 of Ch.2 of Hartman (1997), we have that for every compact set $C\subset \mathbb{R}_{++}$, there exists $t^+\in [0,t^*[$ such that if $t^+<t<t^*$, then $k^*(t)\notin C$. This implies that either $\lim_{t\to t^*}k^*(t)=+\infty$ or $\lim_{t\to t^*}k^*(t)=0$.

Define
\[g(k)=p_2(k-k_2)+f(k_2).\]
Because $f$ is concave, we have that $g(k)\ge f(k)$ for all $k\ge 0$. Consider the differential equation:
\begin{equation}
\dot{k}(t)=g(k(t)),\ k(0)=k.\label{NC2}
\end{equation}
The solution to (\ref{NC2}) is
\begin{equation}\label{NC3}
k^+(t)=e^{p_2t}[k+A(1-e^{-p_2t})],
\end{equation}
where $A=\frac{f(k_2)-p_2k_2}{p_2}\ge 0$. By Lemma 1, $k^*(t)\le k^+(t)$ for all $t\in [0,t^*[$, which implies that $\limsup_{t\to t^*}k^*(t)\le k^+(t^*)$. Hence, we have that $\lim_{t\to t^*}k^*(t)=0$. Choose any $\varepsilon>0$ such that $f(\varepsilon)>0$ and $\varepsilon<k$. Because $f$ is concave and $p_2>0$, there exists such an $\varepsilon$. Define $t^+=\sup\{t\ge 0|k^*(t)\ge \varepsilon\}$. Because $k(0)=k$ and $\lim_{t\to t^*}k^*(t)=0$, $0<t^+<t^*$. Because $f(k^*(t^+))>0$, there exists $\delta>0$ such that $k^*(t)>\varepsilon$ for $t\in [t^+,t^++\delta]$, which is a contradiction. Therefore, the positive nonextendable solution $k^*(t)$ is defined on $\mathbb{R}_+$. Moreover, by the above proof, we have that $k^*(t)\ge \varepsilon$ for all $t\ge 0$. This implies that $t\mapsto \hat{k}(t,k)$ is defined on $\mathbb{R}_+$ and $\inf_{t\ge 0}\hat{k}(t,k)\ge \varepsilon>0$. This completes the proof of our claims.

\subsection{Nondecreasing Property and Concavity of the Value Function}
In this subsection, we show that the value function becomes nondecreasing and concave.

First, choose any $k,k'>0$ such that $k<k'$, and any $(k(t),c(t))\in A_k$. Consider the following differential equation:
\[\dot{k}(t)=f(k(t))-c(t),\ k(0)=k'.\]
By Lemma 1, for any solution $\bar{k}(t)$ to the above equation, $\bar{k}(t)\le \hat{k}(t,k')$. Choose any nonextendable solution $k^*(t)$ to the above equation. If $k^*(t)$ is defined on $\mathbb{R}_+$, then $(k^*(t),c(t))\in A_{k'}$. If $k^*(t)$ is not defined on $\mathbb{R}_+$, then $k^*(t)=0$ for some $t\ge 0$. Define $t^*=\inf\{t\ge 0|k(t)=k^*(t)\}$. Because $k(t)\ge 0$ for all $t$, $t^*$ is well-defined. Define $\bar{k}(t)=k^*(t)$ for $t\in [0,t^*]$ and $\bar{k}(t)=k(t)$ for $t\ge t^*$. Then, $(\bar{k}(t),c(t))\in A_{k'}$. This indicates that $\bar{V}(k)\le \bar{V}(k')$, as desired.

Second, choose any $k_0,k_1>0$, $(k_0(t),c_0(t))\in A_{k_0}$ and $(k_1(t),c_1(t))\in A_{k_1}$. For $s\in [0,1]$, define $k_s=(1-s)k_0+sk_1$, and
\[k_s(t)=(1-s)k_0(t)+sk_1(t),\ c_s(t)=f(k_s(t))-\dot{k}_s(t).\]
Then, $(k_s(t),c_s(t))\in A_{k_s}$. Moreover, $c_s(t)\ge (1-s)c_0(t)+sc_1(t)$ by the concavity of $f$, which implies that
\[\bar{V}(k_s)\ge (1-s)\bar{V}(k_0)+s\bar{V}(k_1),\]
as desired. This completes the proof. $\blacksquare$

\subsection{Proof of Theorem 2}
First, we show that
\begin{equation}\label{GC}
\lim_{t\to \infty}e^{-\rho t}\bar{V}(\hat{k}(t,k))=0
\end{equation}
for every $k>0$. In the proof of the previous subsection, we have already shown that $0<\varepsilon<\hat{k}(t,k)\le k^+(t)$, where $k^+(t)$ is defined in (\ref{NC3}). Because $\bar{V}$ is a concave classical solution to (\ref{HJB}), it is differentiable, and
\[\bar{V}(k')\le \bar{V}(k)+\bar{V}'(k)(k'-k).\]
Therefore,
\[\limsup_{t\to\infty}e^{-\rho t}\bar{V}(\hat{k}(t,k))\le \limsup_{t\to \infty}e^{-\rho t}(\bar{V}(k)+\bar{V}'(k)(k^+(t)-k))=0.\]
On the other hand, because $\bar{V}$ is nondecreasing,
\[\liminf_{t\to\infty}e^{-\rho t}\bar{V}(\hat{k}(t,k))\ge \liminf_{t\to \infty}e^{-\rho t}\bar{V}(\varepsilon)=0,\]
which indicates (\ref{GC}) holds.

Next, consider the ordinary differential equation (\ref{SOL}). Note that, by (\ref{HJB}), if $\bar{V}'(k)=0$, then $\sup_{c\ge 0}u(c)=\rho V(k)$, which is impossible. Therefore, we have that $\bar{V}'(k)>0$ for all $k>0$. Because $u'$ and $\bar{V}'$ are continuous, $(u')^{-1}(\bar{V}'(k))$ is a continuous function. Therefore, by Peano's existence theorem, there exists a continuously differentiable solution $k(t)$ to (\ref{SOL}) defined on $[0,T]$ for some $T>0$.\footnote{See Ch.2 of Hartman (1997).} By Lemma 1, we have that $k(t)\le \hat{k}(t,k)$ for $t\in [0,T]$. Choose any $\varepsilon'>0$ such that $\varepsilon'<\min\{k,k_1\}$. Because $D_+f(k_1)>0$, we have that $f(\varepsilon')>0$. We show that $k(t)\ge \varepsilon'$ for all $t\in [0,T]$. Suppose not. Then, there exists $t^*\in [0,T]$ such that $k(t^*)<\varepsilon'$. Because $k(0)=k>\varepsilon'$, we have that $t^*>0$. By the mean value theorem, we can assume without loss of generality that $t^*<T$ and $\dot{k}(t^*)<0$. Because $\bar{V}$ is concave, by Alexandrov's theorem, it is twice differentiable almost everywhere.\footnote{See Alexandrov (1939) or Howard (1998).} Therefore, we can also assume without loss of generality that $\bar{V}$ is twice differentiable at $k(t^*)$. Define $c(t)=(u')^{-1}(\bar{V}'(k(t)))$. Note that because $\bar{V}$ is a classical solution to (\ref{HJB}),
\[\rho \bar{V}(k(t))=(f(k(t))-c(t))\bar{V}'(k(t))+u(c(t))\]
for all $t\in [0,T]$. Thus, if $h>0$ is sufficiently small, by the usual mean value theorem and the mean value theorem for subdifferentials, there exist $k_3,k_4\in [k(t^*+h),k(t^*)]$, $\theta\in [0,1]$ and $p\in \partial f(k_4)$ such that
\begin{align*}
&~\rho\bar{V}'(k_3)(k(t^*+h)-k(t^*))\\
=&~\rho(\bar{V}(k(t^*+h))-\bar{V}(k(t^*)))\\
=&~(f(k(t^*+h))-c(t^*+h))\bar{V}'(k(t^*+h))-(f(k(t^*))-c(t^*))\bar{V}'(k(t^*))\\
&~+u(c(t^*+h))-u(c(t^*))\\
=&~(f(k(t^*+h))-f(k(t^*)))\bar{V}'(k(t^*+h))-(c(t^*+h)-c(t^*))\bar{V}'(k(t^*+h))\\
~&+(f(k(t^*))-c(t^*))(\bar{V}'(k(t^*+h))-\bar{V}'(k(t^*)))+u(c(t^*+h))-u(c(t^*))\\
=&~p\bar{V}'(k(t^*+h))(k(t^*+h)-k(t^*))+\dot{k}(t^*)(\bar{V}'(k(t^*+h))-\bar{V}'(k(t^*)))\\
&~+(u'(c(t^*+\theta h))-\bar{V}'(k(t^*+h)))(c(t^*+h)-c(t^*)).
\end{align*}
Hence,
\begin{align}
&~\frac{(\rho\bar{V}'(k_3)-p\bar{V}'(k(t^*+h)))(k(t^*+h)-k(t^*))}{h}\nonumber \\
=&~\frac{(u'(c(t^*+\theta h))-\bar{V}'(k(t^*+h)))(c(t^*+h)-c(t^*))}{h}\label{EVALP}\\
&~+\frac{\dot{k}(t^*)(\bar{V}'(k(t^*+h))-\bar{V}'(k(t^*)))}{h}.\nonumber
\end{align}
Because $k_4<\varepsilon'<k_1$, $p\ge D_+f(k_1)>\rho$. Therefore,
\[\liminf_{h\downarrow 0}\frac{(\rho \bar{V}'(k_3)-p\bar{V}'(k(t^*+h)))(k(t^*+h)-k(t^*))}{h}>0.\]
On the other hand, the second term on the right-hand side of the equation (\ref{EVALP}) is always non-positive. Moreover, by the definition of $c(t)$,
\[u'(c(t^*+\theta h))=\bar{V}'(k(t^*+\theta h)).\]
Therefore, we have that the absolute value of the first term of the right-hand side of (\ref{EVALP}) is bounded from
\[\frac{\bar{V}'(k(t^*+h))-\bar{V}'(k(t^*))}{h}|c(t^*+h)-c(t^*)|\to 0\mbox{ as }h\downarrow 0,\]
which leads a contradiction.

Hence, we have that for every solution to (\ref{SOL}) defined on $[0,T]$ for some $T>0$, $\varepsilon'\le k(t)\le \hat{k}(t,k)$ for all $t\in [0,T]$. By Theorem 3.1 of Ch.2 of Hartman (1997), there exists a nonextendable solution $k^*(t)$ to (\ref{SOL}) defined on $\mathbb{R}_+$. Note that, $\varepsilon'\le k^*(t)\le \hat{k}(t,k)$ for all $t\ge 0$.

Define $c^*(t)=(u')^{-1}(\bar{V}'(k^*(t)))$. Note that $\inf_{t\ge 0}c^*(t)\ge (u')^{-1}(\bar{V}'(\varepsilon'))$, and thus
\[\int_0^{\infty}e^{-\rho t}u(c^*(t))dt\]
can be defined, which implies that $(k^*(t),c^*(t))\in A_k$. By (\ref{HJB}),
\[\rho \bar{V}(k^*(t))=\dot{k}^*(t)\bar{V}'(k^*(t))+u(c^*(t)).\]
Therefore,
\begin{align*}
\int_0^Te^{-\rho t}u(c^*(t))dt=&~\int_0^Te^{-\rho t}[\rho \bar{V}(k^*(t))-\dot{k}^*(t)\bar{V}'(k^*(t))]dt\\
=&~\int_0^T-\frac{d}{dt}[e^{-\rho t}\bar{V}(k^*(t))]dt\\
=&~\bar{V}(k)-e^{-\rho T}\bar{V}(k^*(T))\to \bar{V}(k)\mbox{ as }T\to \infty,
\end{align*}
where the last evaluation arises from (\ref{GC}) and $\inf_{t\ge 0}k^*(t)\ge \varepsilon'>0$. Hence,
\[\int_0^{\infty}e^{-\rho t}u(c^*(t))dt=\bar{V}(k),\]
as desired. This completes the proof of Theorem 2. $\blacksquare$

\subsection{Proof of Theorem 3}
Because of Theorem 1, $\bar{V}$ is a concave nondecreasing classical solution to (\ref{HJB}). Moreover, in the proof of Theorem 2, we showed that $\bar{V}$ satisfies (\ref{GC}). Note that, by the proof of Theorem 2, $\bar{V}'(k)>0$ for all $k>0$, and thus $\bar{V}$ is increasing.

Now, suppose that $V$ is any concave increasing classical solution to (\ref{HJB}). Define $g(k)=p_2(k-k_2)+f(k_2)$. Then, $f(k)\le g(k)$ for all $k>0$. Consider the equation (\ref{NC2}). The solution $k^+(t)$ is defined by (\ref{NC3}), and $k^+(t)\ge \hat{k}(t,k)$ for all $t\ge 0$. Because $V$ is concave,
\[V(k)\le V'(k_2)(k-k_2)+V(k_2).\]
Therefore, for some constants $A\ge 0,\ B\in\mathbb{R}$,
\[\limsup_{t\to \infty}e^{-\rho t}V(\hat{k}(t,k))\le \limsup_{t\to\infty}e^{-\rho t}(Ak^+(t)+B)=0.\]
On the other hand, because $\inf_{t\ge 0}\hat{k}(t,k)>0$, 
\[\liminf_{t\to \infty}e^{-\rho t}V(\hat{k}(t,k))\ge 0,\]
which implies that
\begin{equation}\label{GC2}
\lim_{t\to \infty}e^{-\rho t}V(\hat{k}(t,k))=0.
\end{equation}
Consider the following differential equation.
\[\dot{k}(t)=f(k(t))-(u')^{-1}(V'(k(t))),\ k(0)=k.\]
By the same arguments as in the proof of Theorem 2, we have that there exists a solution $k(t)$ defined on $\mathbb{R}_+$ such that $\inf_{t\ge 0}k(t)>0$. Define $c(t)=(u')^{-1}(V'(k(t)))$. Then, by (\ref{HJB}),
\[\rho V(k(t))=\dot{k}(t)V'(k(t))+u(c(t)),\]
and thus, again by the same arguments as in the proof of Theorem 2, we have that
\[\int_0^{\infty}e^{-\rho t}u(c(t))dt=V(k).\]
On the other hand, because $\bar{V}(k)$ is a solution to (\ref{HJB}),
\[\rho \bar{V}(k(t))\ge \dot{k}(t)\bar{V}'(k(t))+u(c(t)).\]
This implies that
\begin{align*}
\int_0^Te^{-\rho t}u(c(t))dt\le&~\int_0^Te^{-\rho t}[\rho \bar{V}(k(t))-\dot{k}(t)\bar{V}'(k(t))]dt\\
=&~\int_0^T-\frac{d}{dt}[e^{-\rho t}\bar{V}(k(t))]dt\\
=&~\bar{V}(k)-e^{-\rho T}\bar{V}(k(T))\to \bar{V}(k)\mbox{ as }T\to 0
\end{align*}
because of (\ref{GC}). Therefore, $V(k)\le \bar{V}(k)$. On the other hand, let $(k^*(t),c^*(t))$ be defined in the proof of Theorem 2. Then, $k^*(t)\ge \varepsilon'$ for all $t\ge 0$. Because $V(k)$ is a solution to (\ref{HJB}),
\[\rho V(k^*(t))\ge \dot{k}^*(t)V'(k^*(t))+u(c^*(t)).\]
This implies that
\begin{align*}
\int_0^Te^{-\rho t}u(c^*(t))dt\le&~\int_0^Te^{-\rho t}[\rho V(k^*(t))-\dot{k}^*(t)V'(k^*(t))]dt\\
=&~\int_0^T-\frac{d}{dt}[e^{-\rho t}V(k^*(t))]dt\\
=&~V(k)-e^{-\rho T}V(k^*(T))\to V(k)\mbox{ as }T\to 0
\end{align*}
because of (\ref{GC2}). Hence, $V(k)\ge \bar{V}(k)$, and thus $\bar{V}\equiv V$. This completes the proof of Theorem 3. $\blacksquare$

\section*{Acknowledgements}
The author thanks to the editor and anonymous referee for their helpful comments and suggestions. This work was supported by JSPS KAKENHI Grant Number JP21K01403.

\section*{References}

\begin{description}
\item{[1]} Acemoglu, D., 2009. Introduction to Modern Economic Growth. Princeton University Press, Princeton.

\item{[2]} Achdou, Y., Buera, F. J., Lasry, J-M., Lions, P-L., Moll, B., 2014. Partial Differential Equation Models in Macroeconomics. Phil. Trans. R. Soc. A 372.

\item{[3]} Alexandrov, A. D., 1939. Almost Everywhere Existence of the Second Differential of a Convex Function and Some Properties of Convex Surfaces Connected with It. Leningrad State Univ. Ann. Math. Ser. 6, 3-35.

\item{[4]} Bardi, M., Capuzzo-Dolcetta, I., 2008. Optimal Control and Viscosity Solutions of Hamilton-Jacobi-Bellman Equations. Birkh\"auser, Boston.

\item{[5]} Barles, G., 1990, An Approach of Deterministic Control Problems with Unbounded Data. Ann. Inst. Henri Poincare 7, 235-258.

\item{[6]} Barro, R. J., Sala-i-Martin, X. I., 2003. Economic Growth. MIT press, Massachusetts.

\item{[7]} Baumeister, J., Leit\~ao, A., Silva, G. N., 2007. The Value Function for Nonautonomous Optimal Control Problems with Infinite Horizon. Syst. Control Lett. 56, 188-196.

\item{[8]} Benveniste, L. M., Scheinkman, J. A., 1979. On the Differentiability of the Value Function in Dynamic Models of Economics. Econometrica, 727-732.

\item{[9]} O. Blanchard, S. Fischer 1989. Lectures on Macroeconomics. MIT press, Massachusetts.

\item{[10]} Cass, D., 1965. Optimum Growth in an Aggregative Model of Accumulation. Rev. Econ. Stud. 32, 233-240.

\item{[11]} Coddington, E. A., Levinson, N., 1984. Theory of Ordinary Differential Equations. McGraw-Hill, New York.

\item{[12]} Crandall, M. G., Lions, P. L., 1983. Viscosity Solutions of the Hamilton-Jacobi Equations. Trans. Am. Math. Soc. 277, 1-42.

\item{[13]} Hartman, P., 1997. Ordinary Differential Equations. 2nd ed. Birkh\"auser Verlag AG, Boston.

\item{[14]} Hermosilla, C., Zidani, H., 2015, Infinite Horizon Problems on Stratifiable State-Constraints Sets. J. Differ. Equations 258, 1430-1460.

\item{[15]} Hermosilla, C., Vinter, R., Zidani, H., 2017, Hamilton-Jacobi-Bellman Equations for Optimal Control Processes with Convex State Constraint, Syst. Control Lett. 109, 30-36.

\item{[16]} Hosoya, Y., 2019. The Magic of Capital. Proceedings of the Tenth International Conference on Nonlinear Analysis and Convex Analysis, 91-108.

\item{[17]} Hosoya, Y., 2023. On the Basis of the Hamilton-Jacobi-Bellman Equation in Economic Dynamics. Physica D 446, 133684.

\item{[18]} Hosoya, Y., 2024. An Elementary Proof of the Euler Equation in Growth Theory, Keio Economic Studies 56.

\item{[19]} Howard, R., 1998. Alexandrov's Theorem on the Second Derivatives of Convex Functions via Rademacher's Theorem on the First Derivatives of Lipschitz Functions. Lecture note from a functional analysis seminar at the University of South Carolina.

\item{[20]} Ioffe, A. D., Tikhomirov, V. M., 1979. Theory of Extremal Problem. North Holland, Amsterdam. 

\item{[21]} Ishii, H. 1989. A Boundary Value Problem of the Dirichlet Type for Hamilton-Jacobi Equations. Annali della Scuola Normale Superiore di Pisa, Classe di Scienze 4e serie, tome 16, 105-135.

\item{[22]} Koopmans, T. C., 1965. On the concept of optimal economic growth. Pontificiae Academiae Scientarum Varia 28, 225-300.

\item{[23]} Lions, P. L., 1982. Generalized Solutions of Hamilton-Jacobi Equations. Pitman Publishing, Massachusetts.

\item{[24]} Malliaris, A. G., Brock, W. A., 1982. Stochastic Methods in Economics and Finance. North Holland, Amsterdam.

\item{[25]} \"Oksendal, B., 2010, Stochastic Differential Equations: An Introduction with Applications, 6th ed. Springer, New York.

\item{[26]} Ramsey, F., 1928. A Mathematical Theory of Saving. Econ. J. 38, 543-559.

\item{[27]} Rockafeller, R. T., 1996. Convex Analysis, Princeton University Press, Princeton.

\item{[28]} Romer, D., 2011. Advanced Macroeconomics. McGraw-Hill, New York.

\item{[29]} Soner, H. M., 1986a. Optimal Control with State-Space Constraint I. Siam J. Control Optim 24, 552-561.

\item{[30]} Soner, H. M., 1986b. Optimal Control with State-Space Constraint II. Siam J. Control Optim 24, 1110-1122.
\end{description}

\end{document}